# On the Reality of "Zero Magnetic" Oscillations of Potential

*Alexander Gritsunov*
Department of Applied Mathematics and Information Technology
Kharkiv National Academy of Municipal Economy
12, Revolution Str., Kharkiv, 61002, Ukraine. E-mail: gritsunov@gmail.com; gritsunov@list.ru

**Abstract:** *The existence of natural Zero Magnetic (ZM) oscillations of the distributed electromagnetic oscillating system (Minkowski space-time) is one of logical consequences of the self-sufficient potential formalism in electromagnetic theory. A mental experiment confirming the reality of the ZM waves of electromagnetic potential is proposed. Main characteristics of those waves are considered. A problem of possibility of application of the ZM waves to information transmission between spatially distributed objects (i.e., radio communication) is formulated.*

### Introduction

Traditionally, the quantum effects are considered as a physical platform for solid-state and optical electronics. Those may be, however, well-promising sources of new ideas also for vacuum electronics. Moreover, some quantum phenomena (e.g., the Aharonov-Bohm effect [1]) are serious causes for reviewing the electromagnetic theory as a whole.

As it is well known, two different ways to describing the electromagnetic phenomena exist: field and potential formalisms. The former is more common but not explaining some recent experiments in the electromagnetism, e.g., the abovementioned Aharonov-Bohm effect. A self-sufficient potential formalism [2, 3] was offered instead treating all electromagnetic phenomena as natural or forced oscillations of some distributed electromagnetic oscillating system (Minkowski space-time). Electromagnetic potential four-vector $\vec{A}^f = \{A_t, A_x, A_y, A_z\}$ components are generalized coordinates of the system describing relative deviation of one from an "undisturbed" state (when both natural and forced oscillations are absent) [4]. This idea makes some positive contribution to the problem of electromagnetic potential reality, which is suspended for a many years.

The self-sufficient potential formalism is more general than the field one. This theory predicts, in particular, existing so-called *Zero Magnetic* (*ZM*) or *Potential* (*P*) electromagnetic potential natural oscillations [2]. Those oscillations have identically zero all components of the "field tensor" [5] (such tensor is interpreted as 4D curl of $\vec{A}^f$ in the potential formalism) and do not transfer energy and pulse.

Therefore, the *ZM* oscillations do not exist from the point of view of the field formalism.

A simplest idealized source of the *ZM* oscillations is the planar capacitor consisting of two symmetrically vibrating charged sheets (*x,y*) with surface densities of charge $-\sigma$ and $+\sigma$. This system emits planar *ZM* waves in $\mp z$ directions with components of the potential

$$A_t^{\mp}(ct,z) = \mp\sqrt{\frac{\mu_0}{\varepsilon_0}}\sigma\varsigma(ct \pm z);$$

$$A_z^{\mp}(ct,z) = +\sqrt{\frac{\mu_0}{\varepsilon_0}}\sigma\varsigma(ct \pm z),$$

where $\mp\varsigma(ct)$ are the *z*-coordinates of the sheets; *c* is the light velocity.

A simplest real emitter of the *ZM* waves is the Hertz dipole. This system radiates those waves in the directions of the dipole axis. However, there are no experimental confirmations of the *ZM* oscillations existence yet.

### Main Part

A mental experiment is proposed to observe the *ZM* waves (see Fig. 1). This is a modification of the Aharonov-Bohm experiment. A short planar electron wave packet $W_1$ falls with velocity $v_{gx} \ll c$ on the diaphragm *D* having two slots. Two charged sheets $C_1$ and $C_2$ with surface densities of charge $-\sigma$ and $+\sigma$ are placed behind the diaphragm transversely to one. The sizes of the sheets in both dimensions ($\Delta X$, $\Delta Y$) are much greater then length of the wave packet $\Delta x$. The sheets are surrounded with an impenetrable for electrons potential barrier *B* of $2\Delta Z \ll \Delta X, \Delta Y$ in width. Initially, the sheets are placed almost together, so electromagnetic potentials the left and the right of ones are practically zero. When parts of the wave packet passed through the slots move close to the middle of the sheets length $\Delta X$ ($W_2$), the sheets are drawn apart for a time of $\Delta t \ll \Delta X / v_{gx}$; then ones are returned to their initial positions. After an interval of order $\Delta Z / c$, components $A_t$ and $A_z$ of the potential four-vector become non-zero for the wave packet part locations (however, the electromagnetic field remains identically equal to zero here). Component $A_z$ does not vary the electron wave phase, while component $A_t$ gives the phase incursion about $-eA_t\Delta t/\hbar$ [6] (*e* is the electron charge). Both parts of the wave packet passed the sheets full length $\Delta X$ ($W_3$) are deflected with the biprism *BP* producing interference figure ($W_4$) on the screen *S*. The *z* coordinates of the interference fringes have to vary depending on were the sheets moved during the wave packet passing or no. Known experimental results of the Aharonov-Bohm effect examinations give a good chance that outcome of the described above mental experiment would be successful.

The fact that the variation of the potential from the relocated sheets $C_1$ and $C_2$ reaches the parts of the electron wave packet $W_2$ only after the time of order $\Delta Z / c$ (not instantaneously) means that some wave process occurs between the sheets and the wave packet parts. This cannot be described from the position of the field formalism, because electromagnetic field of the charged sheets does not affect the electron wave packet at all. This is the *ZM* wave of electromagnetic potential.

A problem of great practical importance is: can the natural *ZM* oscillation be detected with an electron wave packet which does not surround spatially the source of this oscillation? In other words, is it possible to establish information transmission between spatially distributed localized objects (i.e., radio communication) using the *ZM* waves? It is shown in [4] that detection of natural *ZM* oscillations from an external radiator cannot be made with a linear quantum system, as the phase incursion [6]

$$\Delta\varphi = \frac{e}{\hbar} \oint_{L^f} \vec{A}^f(t,x,y,z) \cdot d\vec{l}^{\,f},$$

along a closed loop $L^f$ is identically equal to zero in this case. However, this does not mean that the natural *ZM* oscillations cannot be detected in principle. An additional consideration is needed.

In the case of success, the *ZM* waves may be ideal means for communicating, e.g., with submarines, as those are not absorbed by any material medium relaxing only with distance. The globe and other celestial bodies are also fully transparent for those waves. Therefore, it may be that the failure in the search for extraterrestrial intelligence (SETI) program is caused by receiving the *TEM* waves, not *ZM* ones, whereas the latter may be more suitable for interplanetary information exchange.

**Conclusion**

Oscillations of electromagnetic potential with type of *Zero Magnetic* objectively exist and can be registered experimentally. However, the possibility of using the *ZM* waves to radio communication is unintelligible yet. Additional investigations are needed, mainly, in quantum mechanics and quantum electrodynamics.

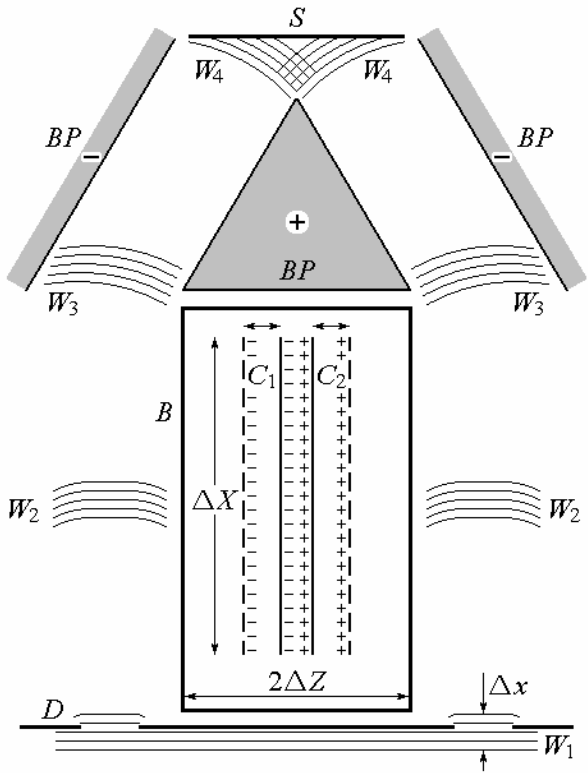

**Figure 1**